\documentclass[twocolumn,showkeywords,ifacmtg]{revtex4}
\usepackage{graphicx}
\usepackage{times}
\usepackage{dcolumn}

\def\tas {$1T$-TaS$_2$}

\begin{document}

\title{Occurrence of  superconductivity when the metal-insulator transition is inhibited in $1T$-TaS${_2}$}
\author{P. Xu, J. O. Piatek, P.-H. Lin, B. Sipos, H. Berger, L. Forr\'o, H. M. R{\o}nnow, and M. Grioni}
\affiliation{Institute of Condensed Matter Physics,
Ecole Polytechnique F\'ed\'erale de Lausanne (EPFL),  CH-1015
Lausanne, Switzerland}
\date{\today}

\begin{abstract}

When a Mott metal-insulator transition is inhibited by a small amount of disorder in the layered dichalcogenide \tas, an inhomogeneous superconducting state arises below T=2.1 K, and coexists with a nearly-commensurate charge-density-wave. By angle-resolved photoelectron spectroscopy (ARPES) we show that it emerges from a bad metal state with strongly damped quasiparticles.  Superconductivity is almost entirely suppressed by an external magnetic field of 0.1 T.

\end{abstract}

\maketitle

One of the most intriguing phenomena in the solid state is the occurrence of electronic  instabilities towards broken symmetry states, like superconductivity (SC), density-waves, charge and orbital order, or the Mott insulating state. When two or more instabilities are simultaneously at work, complex phase diagrams and unusual physical properties are found, and new states of matter may eventually emerge. Understanding such a coexistence or competition is therefore  important. A related open issue is the emergence of SC from a normal state where the nature of the quasiparticles (QPs) is strongly modified by the interactions. Spectroscopic probes of the electronic states with momentum selectivity, namely angle-resolved photoelectron spectroscopy (ARPES) can address these issues.

Broken-symmetry phases are especially prominent in low-dimensional -- quasi one-dimensional (1D) and quasi-two-dimensional (2D) -- materials \cite{Gruner}. The 2D transition metal dichalcogenides (TMDs) in particular, exhibit a variety of  charge-density-waves (CDW) transitions \cite{Wilson}. In their trigonal prismatic (2H) polytypes some of them also support SC, with critical temperatures as high as 7.2 K for NbSe$_2$.
ARPES has been used to identify in the band structure the favorable conditions for electronic instabilities \cite{Straub,Yokoya,Borisenko,Feng}.
According to the simplest Peierls scenario, the CDW is the result of a moderately strong electron-phonon ($e$--$ph$) interaction connecting electron  and hole states across the well-nested Fermi surface (FS). SC is also a consequence of $e$--$ph$ coupling, leading to the formation of Cooper pairs at the FS. Even if alternative scenarios been proposed which do not rely on nesting \cite{Rice, Castro}, both CDW and SC tend to  gap parts of the same FS, and are generally considered to be competing.

SC is usually not found in TMDs of the 1T polytype, with TM ions in octahedral coordination, but it can be induced by doping or by external pressure. Recently, much interest has been raised by the observation that the balance between CDW and SC can be continuously tuned in the TMD TiSe$_2$ by a controlled intercalation of Cu atoms \cite{Morosan}. Here we consider the occurrence of SC in the isostructural TMD \tas, which exhibits a unique sequence of CDW and Mott phases, as a result of the interplay of $e$--$ph$ and Coulomb interactions \cite{Wilson}. At room temperature (RT) the CDW is nearly-commensurate (NC). It consists of an hexagonal array of domains separated by domain walls (discommensurations) where the CDW phase changes rapidly. The domain size and the CDW amplitude grow as temperature is reduced, down to T$_{MI}\sim$180 K, where an insulating commensurate (C) ($\sqrt {13} \times \sqrt {13}$)R 13.9$^\circ$ CDW phase is suddenly established. In real space the Mott metal-insulator (MI) transition corresponds to the localization of one Ta 5$d$ electron at the center of the 13-atom `star of David' cluster unit of the CDW. In reciprocal space, it is due to the opening of a correlation gap in a narrow half-filled band straddling the Fermi level \cite{Tosatti}. Sipos $et~al.$ \cite{Sipos} have shown that the application of external pressure suppresses the Mott transition, and that SC appears above 2.5 GPa. Here we show that, in the presence of a very small amount of disorder which also suppresses the Mott transition, SC arises at ambient pressure below T=2.1 K. We have probed the electronic structure by ARPES, and found that SC emerges from a normal state where  strongly damped quasiparticles bear the spectral signatures of the underlying CDW and of the incipient Mott transition.

It is known that tiny amounts of disorder -- induced e.g. by irradiation or non-stoichiometry -- suppress the MI transition in \tas~\cite{DiSalvo, Mutka, Zwick}. We have taken the systematic approach of producing samples with variable amounts of Cu intercalated between the TaS$_2$ planes, by adding CuCl$_2$ during the crystal growth by chemical-vapor transport.  Here we report data for samples with a Cu content at the detection limit ($<$ 0.1 \%), for which the particular growth conditions yielded a small Ta deficiency and possibly a  small amount of structural disorder.
The stoichiometry was Ta$_{1-x}$S$_2$ with $x=0.015$, as determined by fluorescence microanalysis. X-ray diffraction yielded a $c$-axis lattice parameter (5.90 \AA) identical to that of pure TaS$_2$. In the following, we will refer to these samples as $d$-TaS$_2$.  The in-plane electrical resistivity was measured by the four-probe method between 0.35 K and RT.
The magnetic susceptibility was measured along the $c$-axis in a 0.29~Oe field oscillating at 546~Hz. To achieve a signal from the $ <100~\mu m$ thin sample, five 1.6~mm$^2$ pieces were stacked 1.5~mm apart, fixed with GE varnish into shelves cut in a carbon fibre rod, thus allowing an expelled field volume from the Meissner effect in excess of the sample volume itself.
For ARPES and low-energy electron diffraction (LEED) the crystals were post-cleaved $in$ $situ$ to obtain mirror-like surfaces. We performed ARPES and core-level photoemission measurements with a SPECS PHOIBOS 150 hemispherical analyzer and a high-brightness monochromatized GAMMADATA He lamp (h$\nu$=21.2 eV, 48.4 eV). The energy and momentum resolution were respectively 10 meV and $\pm$ 0.01 \AA$^{-1}$. The Fermi level position was determined with an accuracy of $\pm$1 meV by measuring the metallic cutoff of a polycrystalline Ag film.

\begin{figure}[t!]
\includegraphics[width=3.6in]{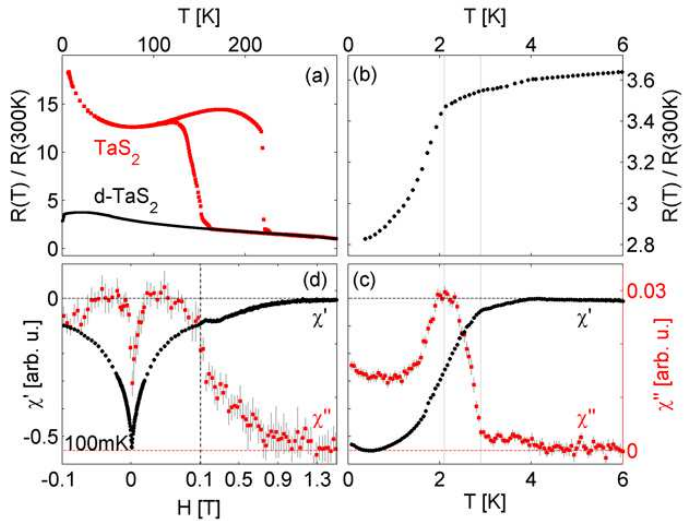}
\caption{(color online) In-plane electrical resistivity and $c$-axis magnetic susceptibility of $d$-TaS$_2$. (a) The resistivity is indistinguishable down to the MI transition at 180-220~K, which is seen in pure TaS$_2$ (red squares) but completely absent in $d$-TaS$_2$ (black circles). (b) The low-temperature part shows a sudden drop in resistivity below 2.1~K, suggesting onset of superconductivity. (c) Temperature dependence of real $\chi'$ (black circles) and imaginary $\chi"$ (red squares) components of the AC-susceptibility measured in zero field. (d) Field dependence of $\chi$ at 100~mK. } \label{fig:1}
\end{figure}

The in-plane electrical resistivity of $d$-TaS$_2$, shown in Fig.\ 1a, increases with decreasing temperature between 300 K and $\sim$20~K. The sharp jump at T$_{MI}$ and the large hysteresis present in the pristine material (TaS$_2$ in short) are completely suppressed in $d$-TaS$_2$. However, above T$_{MI}$ the two curves are indistinguishable, indicating a minimally perturbed sample.
In this temperature range LEED measurements show the coexistence of two equivalent domains corresponding to the NC CDW, rotated in opposite directions. In our experience, cleaved surfaces of pristine \tas~ generally yield single-domain patterns. The overall temperature dependence is similar to that of other doped or disordered TaS$_2$ crystals \cite{DiSalvo, Mutka, Zwick}.
Rather unexpectedly, the resistivity turns over below $\sim$15~K and presents a sharp drop at 2.1~K (Fig.\ 1b), suggesting the onset of superconductivity. The finite residual resistance at 350~mK indicates that superconductivity occur in spatially separated minority regions, which do not percolate through the material.

Indeed, the occurrence of superconductivity is confirmed by the magnetic susceptibility (Fig.\ 1c). The real part $\chi$' becomes negative due to the Meissner effect. The imaginary (dissipative) part $\chi$'' displays a broad peak similar to the behavior observed in granular superconductors. In granular superconductors, the dissipation is linked to junctions between superconducting grains. We speculate that the dissipation in our samples stem from the junctions between superconducting droplets. Such droplets could likely be nucleated by the CDW domain structure. The onset of changes in both $\chi$' and $\chi$'' is at 2.9~K, with the peak in the imaginary part coinciding with the drop in resistivity at 2.1~K.  Susceptibility shows that separate superconducting regions form below 2.9~K, and phase coherence between them begins at 2.1~K, but does not lead to percolation. This picture is supported by the weakness of the Meissner effect testifying a reduced superconducting volume fraction $v$=(V$_{sc}$/V), and is consistent with the field dependence, where an external field of $<0.1$~T suppresses most the Meissner effect, but a portion corresponding to the 2.1-2.9~K signal persist up to 1~T (Fig.\ 1d). Our non-standard set-up prevents a direct, quantitative determination of $v$ from $\chi$', but an estimate is possible from the resistivity jump at the transition. A simple 2D model based on an effective medium approximation, which neglects the smaller conductivity along the $c$ axis, yields $v/v_c=(\Delta R/R)$, where $v_c=0.676$ is the critical value for percolation. In our case we obtain $v=0.15$, well below the percolation limit.

Photoemission core level spectra provide an independent local probe of the CDW state \cite{Pollak, Hughes}. In the presence of a spatial charge modulation, inequivalent Ta atoms experience different local potentials -- corresponding to charge accumulation of depletion --  which translates into differences in the kinetic energies of the photoelectrons. The intrinsic energy width of the shallow Ta 4$f$ core levels is small enough that the resulting spectral broadening or splitting can be easily determined. Figure 2 illustrates the Ta 4$f$ spin-orbit doublet of TaS$_2$ and of $d$-TaS$_2$, measured at the same temperature T=70 K. The spectrum of TaS$_2$ is consistent with literature data for the C phase. Both spin-orbit partners are clearly split, according to the 6:6:1 distribution of inequivalent sites within the 13-atom cluster. For $d$-TaS$_2$ the splitting is smaller, and the individual components cannot be resolved. This is characteristic of the NC phase at T$>$T$_{MI}$, reflecting a CDW amplitude and charge inhomogeneity smaller than in the C phase, and the contribution from domain boundaries. The core level data confirm microscopically that the electronic charge distribution of $d$-TaS$_2$ is different from that of the pristine sample at the same temperature even if, within the domains, the two configurations are quite similar.

\begin{figure}[t!]
\includegraphics[width=2.8in]{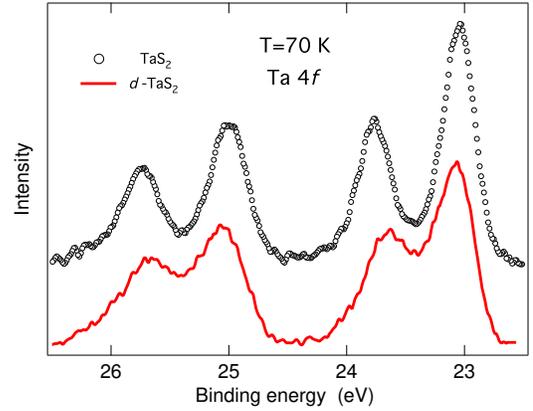}
\caption{(color online) Ta 4$f$ core level spectra (h$\nu$=48.4 eV, HeII$\beta$ line) measured at T=70 K. The broadening and splitting of
the line shapes reflects inequivalent Ta sites in the CDW state. }
\label{fig:2}
\end{figure}

\begin{figure}[h!]
\includegraphics[width=3.2in]{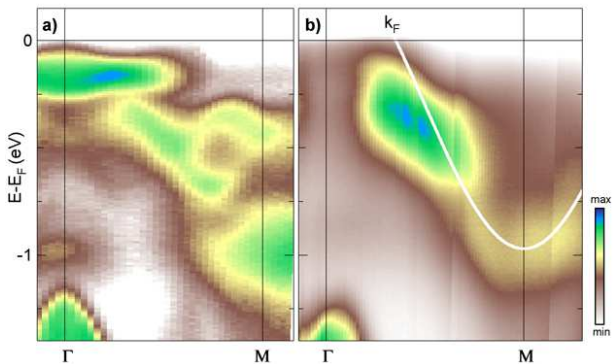}
\caption{(color online) HeI (h$\nu$=21.2 eV) ARPES intensity maps of (a) TaS${_2}$ and (b) $d$-TaS${_2}$ measured at 70 K along the high symmetry $\Gamma$M
direction. The continuous line reproduces a tight-binding band for the unreconstructed structure from Ref. 17, and is a guide to the eye. } \label{fig:3}
\end{figure}

Figure 3 (a) and (b) shows energy-wave vector ARPES intensity maps of TaS${_2}$ and $d$-TaS${_2}$ measured at 70 K along the high symmetry $\Gamma$M direction ($\overline{\Gamma M}$=1.08 \AA$^{-1}$) of the hexagonal Brillouin zone (BZ). The top of the S 3$p$ band is visible at $-$1.3 eV around $\Gamma$. The Ta 5$d$ conduction band disperses upwards from the M point. It is rather diffuse in $d$-TaS${_2}$, and the map shows no clear Fermi level crossing, but a constant-energy cut -- a momentum distribution curve -- at $E$=$E_F$ (not shown) yields a Fermi wave vector $k_F$=0.27 $\overline{\Gamma M}$. Overlayed on the map as a guide to the eye is the result of a tight-binding (TB) calculation for the undistorted structure \cite{Rossnagel}. The superlattice potential splits this band into three separate manifolds. Two of them, centered at $-$1 eV and $-$0.6 eV, are fully occupied. A third half-filled band straddling $E_F$ is further split at T$_{MI}$ with the opening of a correlation gap between an occupied lower Hubbard band (LHB) at $-$0.2 eV and an unoccupied upper band above E$_F$  \cite{Smith, Manzke, Claessen, Aiura}. Clearly the three CDW subbands are not well defined for $d$-TaS$_2$ but the incipient CDW-induced splitting is revealed by strong deviations from the TB dispersion. This is again typical of  the NC phase, where free electrons from the discommensurations can provide enough screening to prevent the Mott transition within the locally commensurate CDW islands. The finite size of the domains, and the superposition of two distinct orientations shown by LEED, contribute to the broad spectral weight distribution.

The effect of disorder on the electronic states is further illustrated in Fig. 4 (a) by a comparison of ARPES data measured at $k$=$k_F$ and T=70 K. The spectra have been normalized to the same integrated intensity over the whole Ta 5$d$ bandwidth, within 1.2 eV of $E_F$. The line shape of TaS$_2$, typical of the insulating C-CDW phase, is dominated by a sharp LHB feature at $-$0.18 eV and has vanishing intensity at $E_F$. The shoulder at $-$0.4 eV is due to the CDW subband. By contrast, the LHB structure is absent in $d$-TaS$_2$. The line shape is not gapped, and exhibits a metallic edge at $E_F$.  

\begin{figure}[h!]
\includegraphics[width=2.5 in]{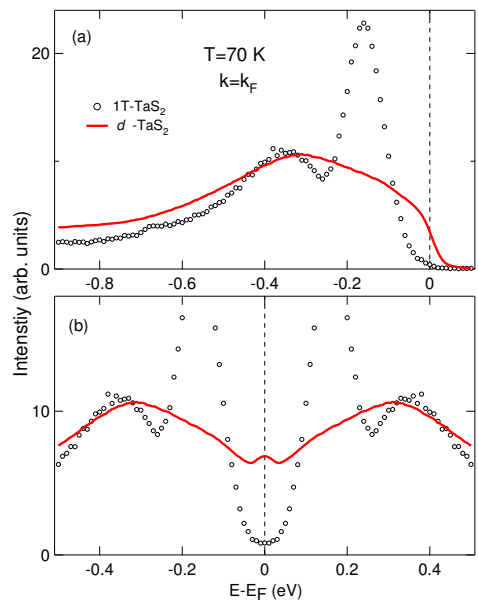}
\caption{(color online) (a) Energy distribution curves measured at $k$=$k_F$ and 70 K from pristine 1T-TaS${_2}$ and
$d$-TaS$_2$.  The same spectra are shown
in (b), after a symmetrization around E$_F$.} \label{fig:3}
\end{figure}

The Fermi cut-off hampers a detailed analysis of the spectral function underlying the ARPES data of Fig. 4 (a). It can be removed by a symmetrization of the spectra around $E_F$: $I^{*}(k_{F}, E)$=$I(k_{F}, E)$ + $I(k_{F}, -E)$. Sufficiently close to $E_F$ the symmetrized spectra shown in Fig. 4 (b) are proportional to the spectral function $A(k_F, E)$, since for the Fermi function $f(E, T)$=$($1$-f(-E, T))$, and  electron-hole symmetry requires $A(k_F, E)$=$A(k_F, -E)$. The line shape of TaS$_2$ is that expected for a Mott insulator. Spectral weight is piled up in the Hubbard subbands, separated by a correlation gap, or  possibly by a deep psuedogap \cite{Georges}. Further away from $E_F$ the loss of electron-hole symmetry may influence the line shape of the upper Hubbard band, but this is irrelevant for the following discussion. The picture of $d$-TaS$_2$ which emerges from Fig. 4 (b) is qualitatively different. $A(k_F, E)$ exhibits a weak coherent QP peak at $E_F$. Spectral weight is removed from $E_F$ and placed in the broad incoherent sidebands. Elsewhere we have presented ARPES data showing the temperature-dependent spectral changes across a Mott transition in the isostructural TMD TaSe$_2$ \cite{PerfettiTaSe}.  The spectrum of $d$-TaS$_2$ closely resembles that of TaSe$_2$ in the metallic side just above the transition. Clearly $d$-TaS$_2$ is on the verge of opening a gap. The broad, intense, sidebands show that the QPs are strongly renormalized both by $e$--$ph$ and Coulomb interactions. 

The electronic properties of TaS$_2$ are complex, due to coexisting interactions and comparable energy scales. The dominant instability is the CDW. At first sight, the Coulomb interaction should be unimportant, since the ratio $\alpha$=$(U/W)$ between the Coulomb correlation energy ($\sim$ 0.4 eV from Fig. 4 (b)) and the conduction bandwidth ($\sim$ 1 eV) is small. In the CDW state, however, the relevant bandwidth is that of the narrower, partly occupied subband at $E_F$. $W$ decreases with temperature as the CDW amplitude increases, and the critical value $\alpha_c$ for the bandwidth-controlled Mott transition is eventually attained. This balance depends on material-specific parameters. In the case of TaSe$_2$ the bulk of the material remains metallic, but a subtle change of parameters yields  $\alpha>\alpha_c$ and a MI transition at the surface. Disorder pins the CDW and prevents the formation of a coherent C phase in $d$-TaS$_2$. In the NC phase, $\alpha<\alpha_c$ : the Coulomb interaction  becomes irrelevant and the MI transition is suppressed.

Favored by the large $e$--$ph$ coupling, a SC state can eventually develop in $d$-TaS$_2$. But what is the nature of the normal state? In the pristine material the sudden disapperence of the Fermi surface at T$_{MI}$ explains the order-of-magnitude increase of the resistivity. $d$-TaS$_2$ retains a small but finite coherent spectral weight at $E_F$ which seems to contradict the insulating-like resistivity. However, the electrical conduction is clearly limited by scattering by the NC domains, and the increasing resistivity most likely reflects the growing amplitude of the CDW. 

A second question concerns the electronic states which pair and give rise to SC. Are these states also involved in the CDW, and eventually in the Mott instabilities? When SC in \tas~ is induced by pressure, there are indications that it develops in regions that are spatially separated from the CDW domains \cite{Sipos}. With increasing pressure the undistorted regions grow at the expense of the commensurate CDW domains, and the critical temperature for SC grows as the CDW is suppressed. A similar scenario may also apply here, but one should be aware of differences between the two physical situations. Pressure modifies the orbital overlap, and therefore the band structure. Disorder, on the other hand, only affects the long-range coherence, leaving the CDW essentially unchanged within the domains, which represent a majority of the sample volume.
We also note that the separation of electronic states within and outside the domains is not so clear-cut. ARPES has shown that, even in the Mott phase, the LHB exhibits a finite ($\sim$ 70 meV) dispersion \cite{Perfetti}. Therefore, the 5$d$ orbital at the star center is not entirely localized, so that states close to the domain boundaries may hybridize with states of appropriate symmetry outside the domain, providing a connection between the two electron systems.

In summary, we have shown that superconductivity can emerge even in the presence of a strong charge-density-wave, when the Mott instability is suppressed in the 2D TMD \tas. ARPES measurements of the normal state show that the coherent weight of the quasiparticles is considerably reduced by the interactions, and that the system remains quite close to a metal-insulator instability. Future experiments will aim at establishing a quantitative relation between the superconducting fraction and the amount of disorder. They should namely provide an answer to the interesting question whether the percolation threshold can be reached before disorder-induced localization sets in.

This work has been supported by the Swiss NSF and by the MaNEP NCCR. We gratefully acknowledge Dr. D. Ariosa for his assistance with the x-ray diffraction measurements.

\end{document}